\documentclass{eas}
\usepackage{graphicx}

\begin{document}

%%-----------------------------
%%      the top matter
%%-----------------------------
\title{Impact of asteroseismology on improving stellar ages determination} 
\runningtitle{Impact of asteroseismology on improving stellar ages determination} 
\author{Y. Lebreton}\address{Observatoire de Paris, GEPI, CNRS UMR 8111, F-92195 Meudon, France and Institut de Physique de Rennes, Universit\'e de Rennes 1, CNRS UMR 6251, F-35042 Rennes, France}
\begin{abstract}
High precision photometry as performed by the CoRoT and Kepler satellites on-board instruments has allowed to detect stellar oscillations over the whole HR diagram. Oscillation frequencies are closely related to stellar interior properties via the density and sound speed profiles, themselves tightly linked with the mass and evolutionary state of stars. Seismic diagnostics performed on stellar internal structure models allow to infer the age and mass of oscillating stars. The accuracy and precision of the age determination depend both on the goodness of the observational parameters (seismic and classical) and on our ability to model a given star properly. They therefore suffer from any misunderstanding of the physical processes at work inside stars (as microscopic physics, transport processes...). In this paper, we recall some seismic diagnostics of stellar age and we illustrate their efficiency in age-dating the CoRoT target HD~52265.
\end{abstract}
\maketitle
%%-----------------------------
%%      your text
%%-----------------------------
\section{Introduction}

While stellar masses and/or radii can be measured directly for some particular stars (members of binary systems, stars observable by interferometry), stellar ages cannot be determined by direct measurements but can only be estimated or inferred. As reviewed by Soderblom (\cite{2010ARAA..48..581S}), there are many methods to estimate the age of a star according to its mass range, evolutionary state and configuration --  single star or star belonging to a group.  For single main sequence (MS) stars, ages are often inferred from empirical indicators (activity or rotation)  and/or from stellar model isochrones which are compared to observed classical parameters  -- usually effective temperature,  luminosity or surface gravity, and metallicity. However the precision and accuracy usually reached are generally not satisfactory (see e.g. Lebreton \& Montalb{\'a}n \cite{2009IAUS..258..419L} and references therein). 

Asteroseismic measurements bring additional constraints on stellar models which allow to greatly improve the age accuracy. Low amplitude solar-like oscillations have now been identified in many stars and error bars on the frequency measurements are currently of a few tenths micro Hertz. Models of the star to age-date can be calculated and adjusted in order to satisfy both the constraints provided by the oscillation frequencies and the classical constraints. Since oscillation modes deeply penetrate inside stars, they provide tools to probe the physical processes at work in stellar interiors, as transport processes, still quite poorly understood. Important by-products of the modelling are the age and mass of stars. 

Here we focus on the determination of the age of  the  {\small CoRoT}  target HD~52265, a solar-like oscillator on the main sequence. The star hosts an exoplanet, the transit of which was not observable but {\small CoRoT} provided a rich solar-like oscillation spectrum that was analysed by Ballot \etal\ (\cite{2011AA...530A..97B}) and modelled recently by Escobar \etal\ (\cite{2012AA...543A..96E}) and Lebreton \& Goupil\ (\cite{2012AA...544L..13L}). In the present study, we concentrate on quantifying the sources of inaccuracy affecting the age of HD~52265. We focus on stellar models inferences, in different contexts where different sets of observational constraints are available and we examine the impact of different possible model input physics on the age.

\section{Observational constraints and seismic diagnostics for age-dating}

\label{obs}

\subsection{Classical data and data extracted from the CoRoT light-curve}

HD~52265 (HIP~33719) is a nearby, metal-rich, single G0V star located at ${{\approx}29}$ pc. To model the star, we considered the following observational data, hereafter called classical data: effective temperature $T_\mathrm{eff}=6116\pm110$~K, metallicity $\mathrm{[Fe/H]}=0.22\pm0.05$ dex, luminosity $L/L_\odot=2.053\pm0.053$ (details are to be given in Lebreton, in preparation).
The spectroscopic surface gravity $\log g=4.32\pm0.20$ dex is not precise enough to be used as a model constraint.

HD~52265 shows a pressure-mode (p-mode) solar-like oscillation spectrum. From the {\small CoRoT} spectrum analysis,  Ballot \etal~(\cite{2011AA...530A..97B}) identified 28 reliable low-degree p-modes of degrees $\ell=0, 1, 2$ and order $n$ in the range $14-24$ (see their Table\ 4). The frequencies $\nu_{n, \ell}$ are in the range $1500-2550\ \mu$Hz with a frequency at maximum amplitude $\nu_\mathrm{max}=2090\pm\  20\ \mu$Hz. The error on each frequency is a few tenths of  $\mu$Hz.  Ballot \etal~also derived a precise value of the rotation period of HD~52265, $P_\mathrm{rot}=12.3{\pm}0.14$ days from the light curve. In the following, the individual frequencies and  $\nu_\mathrm{max}$ are used as model constraints.

\subsection{Derived asteroseismic diagnostics}

In addition to the individual frequencies, we used specific differences of frequencies as diagnostics for stellar models. The large frequency separation $\Delta \nu_{\ell}$, the small frequency separation $d_{02}$ and the Roxburgh \& Vorontsov (\cite{2003AA...411..215R}) separations $dd_{01}$ and $dd_{10}$ are defined as
\begin{eqnarray}
\Delta \nu_{\ell}(n){=}\nu_{n,\ell}{-}\nu_{n-1,\ell}\ ;\  d_{02}(n){=}\nu_{n,0}{-}\nu_{n-1,2}
\end{eqnarray}
\begin{eqnarray}
dd_{01}(n)= \frac{1}{8}(\nu_{n-1, 0}-4\nu_{n-1, 1}+6\nu_{n, 0}-4\nu_{n, 1}+\nu_{n+1, 0})
\end{eqnarray}
\begin{eqnarray}
dd_{10}(n)= -\frac{1}{8}(\nu_{n-1, 1}-4\nu_{n, 0}+6\nu_{n, l}-4\nu_{n+1, 0}+\nu_{n+1, 1}).
\end{eqnarray}
The frequency separation ratios write
\begin{eqnarray}
r_{02}(n)= d_{02}(n)/\Delta \nu_{1}(n)
\end{eqnarray}
\begin{eqnarray}
rr_{01}(n)= dd_{01}(n)/\Delta \nu_{1}(n)\ ;\  rr_{10}(n)= dd_{10}(n)/\Delta \nu_{0}(n+1).
\end{eqnarray}

In absence of rotation and in the adiabatic approximation the frequency of an eigen p-mode of radial order $n$ and degree $\ell$, only depends on two parameters, for instance the density $\rho$ and the adiabatic sound speed $c_\mathrm{s}=(\Gamma_1 P/\rho)^{1/2}$ ($\Gamma_1$ and $P$ are the first adiabatic index and the pressure, respectively). For a perfect gas, $c_\mathrm{s}\propto(T/\mu)^\frac{1}{2}$ where $T$ is the temperature and $\mu$ the mean molecular weight.
Following the asymptotic approximation (Tassoul\ \cite{1980ApJS...43..469T}),  for high $n$ and $\ell{\ll}n$,  $\nu_{n,\ell}$ can be written
\begin{eqnarray}
\label{asym}
\nu_{n,\ell}{=}\Delta \nu\left( n+\frac{1}{2}\ell +\epsilon\right)-\ell(\ell+1) D_0
\end{eqnarray}

\begin{figure}[!ht]
\begin{center}
\resizebox{0.6\hsize}{!}{\includegraphics{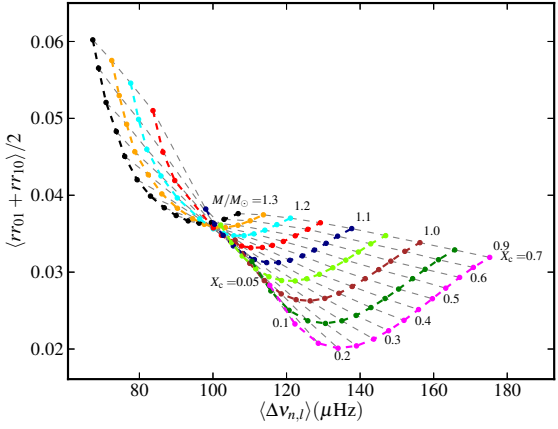}}
\caption{Asteroseismic diagram showing the run of $(\langle rr_{\mathrm{01}}\rangle+ \langle rr_{\mathrm{10}}\rangle)/2$ as a function of $\langle \Delta \nu_{\ell}(n)\rangle$ for stars with masses in the range $0.9-1.3\ M_\odot$ during the MS (the central H abundances $X_\mathrm{c}$ are pinpointed). Models have an initial He abundance $Y=0.275$, metallicity $Z/X=0.0245$ and mixing-length parameter $\alpha_\mathrm{conv}=0.60$.}
\label{astdiag}
\end{center}
\end{figure}

The quantity $\epsilon$ is sensitive to the physics of surface layers but weakly sensitive to $n$ and $\ell$, while $D_0$ is sensitive to the interior sound speed gradient. The large separation $\Delta \nu_{\ell}(n){\equiv}\Delta \nu$ is approximately constant whatever the $\ell$ value while the small separation $d_{02}$ is $d_{02} {\approx}6D_0$. The large separation measures the inverse of the sound travel time across a stellar diameter which is related to the mean density. The small separation $d_{02}$  probes the evolution status (then age) of the star through the sound speed gradient built by the changes of mean molecular weight in the inner regions. Ulrich\ (\cite{1986ApJ...306L..37U}) and Christensen-Dalsgaard\ (\cite{1988IAUS..123..295C}) proposed to use the couple ($\langle\Delta\nu\rangle$, $\langle d_{02}\rangle$) as a diagnostic of age and mass on the MS. However, these quantities are much sensitive to the badly understood near-surface physics. Later on,  Roxburgh \& Vorontsov (\cite{2003AA...411..215R}) demonstrated that the separation ratios $r_{02}$ and $rr_{01/10}$ are quite insensitive to near-surface effects so that Ot{\'{\i}} Floranes \etal\ (\cite{2005MNRAS.356..671O}) used the couple ($\langle\Delta\nu\rangle$, $\langle r_{02}\rangle$) as age and mass diagnostic. The mean values of the ratios $rr_{01/10}$ also probe stellar cores and are sensitive to age. This is illustrated in Fig.~\ref{astdiag} that shows the run of $(\langle rr_{\mathrm{01}}\rangle+ \langle rr_{\mathrm{10}}\rangle)/2$ as a function of $\langle \Delta \nu\rangle$ along the MS of stars of different masses. For all masses, the $\langle rr_{01/10} \rangle$ ratio decreases at the beginning of the evolution on the MS down to a minimum and then increases up to the end of the MS. The minimum value is larger and occurs earlier as the mass of the star increases, i.e. as a convective core appears and develops.

As for HD~52265, from the observed individual oscillation frequencies, we derived  $\langle\Delta\nu_{\ell}(n)\rangle=98.13\pm0.14\ \mu$Hz, $\langle d_{02}(n)\rangle= 8.20\pm 0.31\ \mu$Hz,  $\langle rr_{02}(n)\rangle=0.0835\pm0.0033$,  $\langle rr_{01}(n)\rangle=0.0331\pm0.0015$ and $\langle rr_{10}(n)\rangle=0.0329\pm0.0016$. These quantities will be used as constraints for the modelling.

\section{Internal structure models of HD~52265 and their oscillations frequencies}
\label{LM}

\subsection{Model and frequencies calculation}

We have modelled the star with the stellar evolution code Cesam2k (Morel \& Lebreton, \cite{2008ApSS.316...61M}) with the aim to evaluate the effect of the choice of the models input physics on the inferred age of HD~52265. We considered the following input physics and parameters (see also Table~\ref{modelinputs}), a full description will be given in Lebreton, in preparation:

\begin{itemize}
\item {\em Opacities, equation of state ({\small EoS}), nuclear reaction rates and microscopic diffusion:}\ As a reference, we used {\small OPAL05 EoS}  (Rogers \& Nayfonov, \cite{2002ApJ...576.1064R}),  {\small OPAL96} and {\small WICHITA} opacities, the latter at low temperatures (Iglesias \etal, \cite{1996ApJ...464..943I}, Fergusson \etal, \cite{2005ApJ...623..585F}), {\small NACRE} nuclear reaction rates (Angulo \etal, \cite{1999NuPhA.656....3A}) except for the $^{14}N(p,\gamma)^{15}O$ rate taken from {\small LUNA} (Formicola \etal, \cite{2004PhLB..591...61F}). Microscopic diffusion of He and heavy elements is included according to Michaud \& Proffitt (\cite{MP93}). For uncertainty estimation we also calculated models based on the {\small OPAL01 EoS}, the {\small NACRE} reaction rate for  $^{14}N(p,\gamma)^{15}O$ and models without diffusion.

\item {\em Convection and overshooting:} We used the {\small CGM} convection theory (Canuto \etal\ \cite{1996ApJ...473..550C}) as the reference and the {\small MLT}  (B{\"o}hmVitense, \cite{1958ZA.....46..108B}) as an alternative. The mixing-length parameter $\alpha_{\rm conv}$ was either derived from the optimization of the models (as in Cases 2-7 in Table~\ref{cases}, see Section~\ref{modcase} below) or fixed to the solar value $\alpha_{\rm conv}$ (Case 1 in Table~\ref{cases}). We fixed $\alpha_{\rm conv}=0.688$ ({\small CGM}) or $1.762$ ({\small MLT}) as was obtained through the calibration of the radius and luminosity of a solar model (see e.g. Morel \& Lebreton\ \cite{2008ApSS.316...61M}). Reference models do not include overshooting. In alternate models, we set the core overshooting distance to be $\ell_{\rm ov, c}{=}\alpha_{\rm{ov}}\times \min(R_{\rm cc}, H_{\rm p})$ ($\alpha_{\rm ov}$ and $R_{\rm cc}$ are the overshooting parameter and the radius of the convective core respectively). In other models, we adopted the Roxburgh\ (\cite{1992AA...266..291R}) prescription, in which overshooting extends on a fraction of the mass of the convective core $M_{\rm cc}$, the mass of the mixed core being expressed as $M_{\rm ov, c}{=}\alpha_{\rm{ov}}\times{M_{\rm cc}}$ with $\alpha_{\rm{ov}}=1.8$. 

\item {\em Atmospheric boundary condition:} We used grey atmospheres (Eddington T-$\tau$ law) as the reference and the Kurucz (\cite{1993yCat.6039....0K}) T-$\tau$ law as an alternative.

\item {\em Solar mixture and stellar chemical composition:} The mass fractions of H, He and heavy elements are denoted by $X$, $Y$ and $Z$ respectively. We used the {\small GN93} mixture  (Grevesse \& Noels, \cite{1993oee..conf...15G}) as the reference and the {\small AGSS09} mixture (Asplund \etal, \cite{2009ARAA..47..481A}) as an alternative. The photospheric $(Z/X)_\odot$ ratio is either $0.0244$ ({\small GN93}) or  $0.0181$  ({\small AGSS09}). The present $(Z/X)$ ratio of HD~52265 is related to the observed $\mathrm{[Fe/H]}$ through $\mathrm{[Fe/H]}=\log(Z/X)-\log(Z/X)_\odot$. The initial $(Z/X)_0$ ratio is derived from model calibration as explained in the following. As for the initial helium abundance $Y_0$ we considered different cases. Either $Y_0$ can be inferred from the model calibration if enough observational constraints are available (Case 3-6 in Table~\ref{cases}) or it has to be fixed (Cases 1-2 in Table~\ref{cases}). In that later case, we either assumed a value for $Y_0$ or we derived it from the helium to metal enrichment ratio $(Y_0 - Y_\mathrm{P})/{(Z- Z_\mathrm{P})}{=}{\Delta Y}/{\Delta Z}$ where $Y_\mathrm{P}$ and $Z_\mathrm{P}$ are the primordial abundances. We adopted $Y_\mathrm{P}{=}0.245$  (Peimbert \etal, \cite{2007ApJ...666..636P}), $Z_\mathrm{P}{=}0.$ and, ${\Delta Y}/{\Delta Z}{\approx}2$ from a solar model calibration in luminosity and radius. 

\end{itemize}

The calculation of the oscillation frequencies was performed using the Belgium {\small LOSC} adiabatic oscillation code (Scuflaire \etal, \cite{2008ApSS.316..149S}).  The values of the individual frequencies are highly dependent on the physics of the external stellar layers. The poorness of the physical description of convection in stellar atmospheres makes the models quite unreliable in these zones and consequently the associated frequencies. These so-called near-surface effects produce an offset between the observed and computed oscillation frequencies. In some models (see below), we applied to the model frequencies the empirical corrections obtained from the seismic solar model by Kjeldsen \etal\ (\cite{2008ApJ...683L.175K}) following the procedure described by Brand{\~a}o \etal\ (\cite{2011AA...527A..37B}).

%%%%%%%%%%%%%%%%%%%%%%%%%%%%%%%%%%%%%%%%%%%%%%%%%%%%%%%%%%%%%%%
\begin{table}  %[t]
\caption{Summary of the different sets of input physics considered for the modelling of HD~52265. As detailed in the text, the reference set of inputs denoted by REF is based on OPAL05 EoS,  OPAL96/WICHITA opacities, NACRE+LUNA  reaction rates (this latter only for $^{14}N(p,\gamma)^{15}O$), the CGM formalism for convection, the MP93 formalism for microscopic diffusion, the Eddington grey atmosphere and GN93 solar mixture and includes neither overshooting, nor convective penetration or rotation. For the other cases we only indicated the input that is changed with respect to the reference.}
\begin{tabular}{lll}
\hline\hline
Set & Input physics  \\
\hline
A & REF   \\
B & convection MLT  \\
C & AGSS09 mixture \\
D & NACRE for $^{14}N(p,\gamma)^{15}O$ \\
E & no microscopic diffusion  \\
F  &Kurucz model atmosphere \\%
G & EoS OPAL01   \\
H   & overshooting $\alpha_{\rm ov}{=}0.15 H_p$   \\%
I   & overshooting $M_{\rm ov, c}{=}1.8\times{M_{\rm cc}}$   \\%
\hline\end{tabular}
\label{modelinputs}
\end{table}
%%%%%%%%%%%%%%%%%%%%%%%%%%%%%%%%%%%%%%%%%%%%%%%%%%%%%%%%%%%%%%%
%%%%%%%%%%%%%%%%%%%%%%%%%%%%%%%%%%%%%%%%%%%%%%%%%%%%%%%%%%%%%%%
\begin{table*}  %[t]
\caption{Summary of the different cases considered for the modelling of HD~52265. Column ``Observed'' lists the constraints considered in the modelling and Column ``Adjusted'' lists the model parameters that can be adjusted.  The letters A and M stand for age and mass respectively.}
\label{cases}
\begin{tabular}{llll}
\hline\hline
Case & Observed  &Adjusted  & Fixed \\
\hline
$1$                     &  $T_{\mathrm{eff}}$, $L$, [Fe/H]                                                 & A, M, $(Z/X)_0$ & $\alpha_{\rm conv}$, $Y_0$\\
$2$ &  $T_{\mathrm{eff}}$, $L$, [Fe/H], $\langle\Delta\nu\rangle$ & A, M, $(Z/X)_0$, $\alpha_{\rm conv}$ & $Y_0$\\
$3$  &  $T_{\mathrm{eff}}$, $L$, [Fe/H], $\langle\Delta\nu\rangle$, $\nu_\mathrm{max}$ & A, M, $(Z/X)_0$, $\alpha_{\rm conv}$, $Y_0$ & --\\
$4$      &  $T_{\mathrm{eff}}$, $L$, [Fe/H], $\langle\Delta\nu\rangle$, $\langle d_{02}\rangle$  & A, M, $(Z/X)_0$, $\alpha_{\rm conv}$, $Y_0$ & --\\
%$5$      &  $T_{\mathrm{eff}}$, $L$, [Fe/H], $\langle rr_{02}\rangle$, $\langle rr_{01/10}\rangle$   & A, M, $(Z/X)_0$, $\alpha_{\rm conv}$, $Y_0$ & --\\
$5$                    &  $T_{\mathrm{eff}}$, $L$, [Fe/H], $r_{02}(n)$, $rr_{01/10}(n)$ & A, M, $(Z/X)_0$, $\alpha_{\rm conv}$, $Y_0$ & --\\
$6$                     &  $T_{\mathrm{eff}}$, $L$, [Fe/H], $\nu_{n, \ell}$                     & A, M, $(Z/X)_0$, $\alpha_{\rm conv}$, $Y_0$ & --\\
\hline
\end{tabular}
\end{table*}
%%%%%%%%%%%%%%%%%%%%%%%%%%%%%%%%%%%%%%%%%%%%%%%%%%%%%%%%%%%%%%%

\subsection{Model calibration}

We have used the Levenberg-Marquardt minimization method in the way described by Miglio \& Montalb{\'a}n  (\cite{2005AA...441..615M}) to adjust the unknown parameters of the modelling (cf Table~\ref{cases}) so that the models of HD~52265 fit at best the observations, within the error bars.  The goodness of the fit is evaluated through the $\chi^2$-minimization:
\begin{eqnarray}
\label{chi2}
%\chi^2 = \frac{1}{N_\mathrm{obs}-1} \cdot\sum_{i=1}^{N_\mathrm{obs}} \frac{\left ( x_\mathrm{i, mod} - x_\mathrm{i, obs} \right ) ^2}{\sigma_\mathrm{i, obs}^2}
\chi^2 = \sum_{i=1}^{N_\mathrm{obs}} \frac{\left ( x_\mathrm{i, mod} - x_\mathrm{i, obs} \right ) ^2}{\sigma_\mathrm{i, obs}^2} 
\end{eqnarray}
where $x_\mathrm{i, mod}$ and  $x_\mathrm{i, obs}$ are the modelled and observed values of the $i^\mathrm{th}$ parameter, respectively. 
The more observational constraints available, the more free parameters can be adjusted in the modelling process. When too few observational constraints are available, some free parameters have to be fixed ---more or less arbitrarily by the modeller as discussed below.

\subsection{Different modeling cases}
\label{modcase}
We considered various cases where $N_\mathrm{par}$ unknown parameters of a stellar model are adjusted to fit $N_\mathrm{obs}$ observational constraints (see Table~\ref{cases}). For each case we made several model optimizations with different sets of input physics as explicated in Table~\ref{modelinputs}. 

\begin{itemize} 
\item {\em Case $1$: Age and mass from global parameters.} In this case solely the classical parameters are constrained by observation ($T_{\mathrm{eff}}$, present [Fe/H], $L$). We therefore sought which mass $M$, age $A$ and initial metal to hydrogen ratio $(Z/X)_0$ are required for the model to satisfy these constraints. Since only 3 unknowns can be adjusted with the 3 observed parameters we had to fix, the other inputs of the models, mainly the initial helium abundance $Y_0$, the mixing-length $\alpha_{\rm conv}$ and overshooting parameter $\alpha_{\rm ov}$. As a reference, we assumed that $Y_0$ can be derived from the helium to metal enrichment ratio ${\Delta Y}/{\Delta Z}=2$. We took $\alpha_{\rm conv, \odot}{=}0.688$ from the solar model calibration and $\alpha_{\rm ov}{=}0.0$. Alternate models consider extreme values of $Y_0$ (the primordial, minimum allowed value, $0.245$), of $\alpha_{\rm conv}$ ($0.550, 0.826$, i.e. a change in $\alpha_{\rm conv, \odot}$ by 20 per cents) and of $\alpha_{\rm ov}$ ($0.15, 0.30$) and different input physics (see Table~\ref{modelinputs}).

\item {\em Cases $2$: Age and mass from large frequency separation $\langle \Delta \nu \rangle$ and classical parameters.} In this case only $\langle \Delta \nu \rangle$  is used as a constraint together with the classical parameters. This would be the situation with the future {\small TESS/NASA} mission. We could then adjust the mixing-length parameter, together with the mass, age and $(Z/X)_0$ (4 unknowns, 4 constraints). We had to fix the initial helium abundance from the helium to metal enrichment ratio. We estimated $\langle \Delta \nu \rangle$  from a detailed frequency calculation and corrected the model frequencies from near-surface effects. 

\item {\em Case $3$: Age and mass from scaled values of $\langle \Delta \nu \rangle$ and $\nu_\mathrm{max}$, and classical parameters.} In this case, the seismic constraints are $\langle \Delta \nu \rangle$ and $\nu_\mathrm{max}$. They are calculated according to the scaling relation of Kjeldsen \& Bedding, \cite{1995AA...293...87K},
i.e.  $$\langle \Delta \nu \rangle/134.9{=}(M/M_\odot)^{1/2} (R/R_\odot)^{-3/2},$$ $$\mathrm{and}\ \nu_\mathrm{max}/3.05{=}(M/M_\odot)(T_\mathrm{eff}/5777)^{-1/2} (R/R_\odot)^{-2}.$$ Frequencies do not need to be explicitly calculated in this case.

\item {\em Case $4$: Age and mass from large frequency separation $\langle \Delta \nu \rangle$, small frequency separation  $d_{02}$  and classical parameters.} In this case the frequencies are explicitly calculated and corrected from near-surface effects. 

\item {\em Cases $5$: Age and mass from frequency separations ratios --$r_{02}$, $rr_{01/10}$-- and classical parameters.}  The use of frequency separation ratios allows to minimise the impact of near-surface effects. The models are constrained by the individual frequency separation ratios $r_{02}(n)$, $rr_{01/10}(n)$ as derived from the individual frequencies. We did not apply  near-surface corrections to the model frequencies. We point out that the model separation ratios and the observed ones are calculated in a consistent way.

\item {\em Case $6$: Age and mass from individual frequencies $\nu_{n, \ell}$ and classical parameters.} We considered the full set of $28$ frequencies as model constraints and corrected the model frequencies from near-surface effects.

\end{itemize}

\section{Results: age from stellar modelling}

\begin{figure}
\begin{center}
      \resizebox{0.7\hsize}{!}
	     {\includegraphics{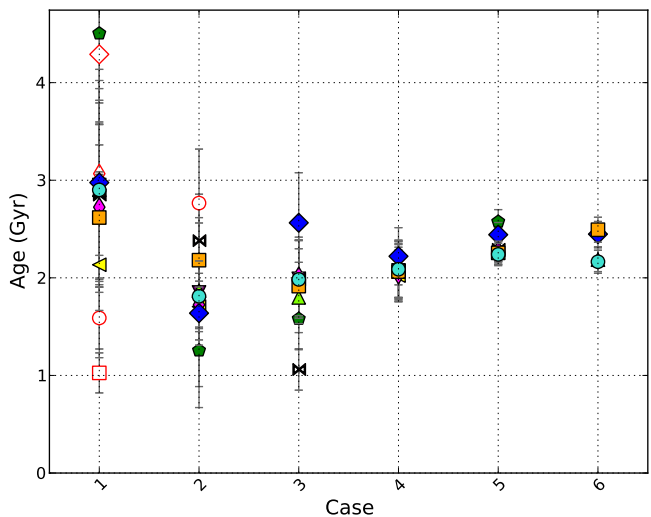}}
	       \caption{Age of HD~52265 inferred from stellar modelling for several sets of observational constraints (see Table~\ref{cases}) and different input physics and free parameters in the models (Table~\ref{modelinputs}). Different symbols are used for the different model specifications listed in Table~\ref{cases}:  turquoise squares (reference models A), orange circles (models B), blue diamonds (models C), small magenta diamonds (models D), green pentagons (models E), brown bow-ties (models F), chartreuse upper triangles (models G), purple down triangles (models H), right yellow triangles (models I). In addition, for models A, are shown the effects of a reasonable change of $\alpha_{\rm conv}$ by $+0.2$ (open red diamond) or $-0.2$ (open red square), of an extreme value of $\alpha_{\rm ov}=0.3$ (open small red diamond)  while open circles are for models with different fixed $Y_0$ values ($Y_0=0.25$, in Case 1 and 0.27, Case 2).
    }
\label{Allages}
\end{center}
\end{figure}

Here, we only discuss in details the results concerning the age-dating of HD~52265. Detailed results of all these models will be published in a forthcoming paper, in what concerns in particular the mass, radius, initial helium abundance, surface gravity of the star. Figure~\ref{Allages} shows the age range obtained for the different cases listed in Table~\ref{cases} considering different sets of input physics and free parameters. 

As can be seen from Column 1 (Case $1$), in Fig.~\ref{Allages}, there is a large scatter ($A=0.8-5.6$ Gyr) in the ages of HD~52265 obtained with different sets of input physics when no seismic observations are available. This is the usual situation of age-dating from classical parameters $L$, $T_\mathrm{eff}$ and [Fe/H]. It is worth to point out that if the error bars on the classical parameters were to be reduced, the error bar on an individual age determination with a given set of input physics would be reduced but the scatter associated to the use of different input physics would remain the same. Note that in that case the values of $Y$, $\alpha_\mathrm{mlt}$ and $\alpha_\mathrm{ov}$ had to be fixed and that different ages result from different choices. In particular a change of  $\alpha_\mathrm{mlt}$ by $\pm 0.2$ around the solar calibrated value induces a change of age of  $\sim 40$ per cents.
In Case $2$ where the large frequency separation is included as a model constraint, the age scatter remains large ($A=0.7-3.3$ Gyr) as expected due to the rather weak sensitivity of $\langle \Delta \nu \rangle$ to age. In Case $3$ where scaling-laws are used to derive $\langle \Delta \nu \rangle$ and $\nu_\mathrm{max}$, the age scatter is $A=0.8-3.1$ Gyr. In Case $4$ the possible age range is narrowed ($A=1.7-2.5$ Gyr) due to the fact that $d_{02}$ is a good indicator of the stellar evolutionary stage. The age is further improved in Case $5$  ($A=2.1-2.7$ Gyr) when the separation ratios $r_{02}(n)$, $rr_{01/10}(n)$ constrain the models. Finally, in Case $6$, as expected, the range of ages is narrowed  ($A=2.0-2.6$ Gyr) when the individual frequencies are included as model constraints. However it is worth noticing that although Cases $5$ and $6$ reach the same precision on age, in Case $6$  the frequencies were corrected from near-surface effects which adds an uncertainty to the models.

To summarize, the detailed modelling of HD~52265 considering both the classical observational constraints and appropriate seismic constraints, mainly the frequency separation ratios, allows to attribute the star an age $A=2.3\pm 0.3$ Gyr. The models also provide the mass $M=1.24\pm 0.04\ M_\odot$ and the radius $1.31\pm 0.04\ R_\odot$. The derived seismic surface gravity is $\log g = 4.30 \pm 0.05$ dex. For HD~52265, the seismic $\log g$ is very close to the spectroscopic one but the error bar is much smaller (0.05 dex vs. 0.20 dex). These results will be discussed with more details in a forthcoming paper.

\section{Conclusions}
This work has aimed at evaluating the ability we have to estimate the age of single stars (that maybe exoplanet hosts) for which asteroseismic observations are available. 
We have first pointed out that one has to be very careful when giving an age estimate even when seismic constraints are considered. Indeed, even if for a given set of input physics and parameters the age determination is quite good, the uncertainties in the input physics and free parameters of the models seriously hamper the age-dating. Furthermore, from the full modelling of HD~2265, we showed that well-chosen asteroseismic constraints really allow to improve the accuracy on age. 
Today with the current error bars on the determination of the oscillation frequencies (about $0.2 \mu$Hz) and on classical parameters, it is possible to estimate the age of a single star to better than $15$ per cents. This is possible only when individual frequencies are available. The best way to proceed to age-date the considered star is to use the frequency separation ratios, if available because they are indicators of the evolutionary stage reached by the star. Further improvement will only come from a better knowledge of the input physics at work in stellar interiors and atmospheres. This will demand even more precise observational constraints and concerns both classical and seismic constraints. In that respect the Gaia mission (to be launched at the end of 2013) and the PLATO mission (still to be selected) will allow great leaps forward. It is only at that price that we should be able to discriminate between models differing in their input physics and therefore  determine their age securely.

%%-----------------------------
%%      your bibliography
%%-----------------------------

\end{document}